\documentclass[prb,aps,floats,twocolumn,showpacs,groupedaddress,floatfix,amsmath]{revtex4}

\def\bc{\begin{center}}
\def\ec{\end{center}}
\def\beq{\begin{equation}}
\def\eeq{\end{equation}}
\def\bw{\begin{widetext}}
\def\ew{\end{widetext}}
\def\bea{\begin{eqnarray}}
\def\eea{\end{eqnarray}}
\def\non{\nonumber}

\def\dag{\dagger}

\def\ra{\rangle}

\def\al{\alpha}
\def\be{\beta}
\def\de{\delta}
\def\ep{\epsilon}

\def\tb{\tilde b}
\def\tv{\tilde v}

\def\pa{\partial}

\renewcommand{\vec}[1]{\mbox{\boldmath$#1$}}
\usepackage{amssymb}
\usepackage{graphicx}
\usepackage{dcolumn}
\usepackage{bm}
\usepackage{subfigure}

\begin{document}

\title{Fractional quantum Hall edge: Effect of nonlinear dispersion and edge roton}
\author{Shivakumar Jolad$^1$, Diptiman Sen$^2$, and Jainendra K. Jain$^1$}
\affiliation{$^1$Department of Physics, Pennsylvania State University, 
University Park, PA 16802}
\affiliation{$^2$Center for High Energy Physics, Indian Institute of Science, 
Bangalore 560012, India}

\date{\today}

\begin{abstract}
According to Wen's theory, a universal behavior of the fractional quantum Hall edge is expected at sufficiently low energies, where the dispersion of the elementary edge excitation is linear. A microscopic calculation shows that the actual dispersion is indeed linear at low energies, but deviates from linearity beyond certain energy, and also exhibits an ``edge roton minimum." We determine the edge exponent from a microscopic approach, and find that the nonlinearity of the dispersion makes a surprisingly small correction to the edge exponent even at energies higher than the roton energy. We explain this insensitivity as arising from the fact that the energy at maximum spectral weight continues to show an almost linear behavior up to fairly high energies. We also formulate an effective field theory to describe the behavior of a reconstructed edge, taking into account multiple edge modes. Experimental consequences are discussed.
\end{abstract}

\maketitle

\section{Introduction}

The edge of a fractional 
quantum Hall (FQH) system [\onlinecite{Tsui}] constitutes a realization of a chiral Tomonaga-Luttinger liquid (CTLL). (The word chiral implies that all the fermions move
in the same direction). In a seminal work, 
Wen postulated that the CTLL at the FQH edge is very special in that the exponent characterizing its long-distance, low-energy physics is a universal quantized number, which depends only on the quantized 
Hall conductance of the bulk state but not on other details
[\onlinecite{WenIntJModPhy,wen90}]. He described the FQH edge through an 
effective field theory approach (EFTA) based on the postulate 
that the electron operator at the edge of the $\nu=1/m$ FQH state has the form
\beq \hat\psi(x) \sim e^{-i\sqrt{m}\hat\phi(x)} \label{Wenansatz} \eeq
where $\hat\phi(x)$ is the bosonic field operator.
The imposition of antisymmetry forces $m$ to be an odd integer 
[\onlinecite{WenIntJModPhy}], which, in turn, leads to quantized 
exponents for various correlation functions.
In particular, it predicts a relation $I\sim V^3$ between the current ($I$) and the voltage ($V$) for tunneling from a three-dimensional Fermi liquid into the 1/3 FQH edge, which has been tested experimentally by Chang {\em et al.} and Grayson {\em et al.} [\onlinecite{Grayson,Chang1,Chang2,Chang3}].

Wen's theory describes the FQHE edge in the asymptotic limit of low energies and long distances. The CTLL description is inapplicable at energies comparable
to or larger than the bulk gap, where bulk excitations become available; we will not consider such high energies in this work. However, even in a range of energies below the bulk gap, deviations from the ideal asymptotic behavior may arise because the dispersion of the elementary edge excitation deviates from linearity and also exhibits an ``edge roton minimum." The aim of this paper is to estimate these corrections from a microscopic approach. To focus on these corrections, we make appropriate approximations (mainly a neglect of composite fermion $\Lambda$ level mixing, discussed previously [\onlinecite{MandalJain}]) that guarantee an ideal quantized behavior at very low energies. 

A deviation from linearity in the dispersion of the elementary edge excitation is expected to produce corrections for the following reason.
In the bosonic model of edge excitations, the spectral weights of all excitations at a given momentum obey a sum rule (see Eq. (\ref{SSWGenl})), first demonstrated by Palacios and MacDonald [\onlinecite{PalaciosMacDonald}], which is valid up to a unitary rotation of the basis. The long time behavior of the Green function and the differential conductance for tunneling from an external Fermi liquid into the FQH edge, on the other hand, are sensitive to the states within an energy slice. 
However, for linear dispersion, the energy and momentum are uniquely related, so the sum rule is also valid for all states at a given {\em energy}, which produces a quantized power law exponent for the differential conductance. (A more detailed discussion is given in Appendix B). A nonlinearity in the dispersion, on the other hand, produces an energy band for excitations, as shown, for example, in Fig. \ref{DispersionFits} below. In the absence of a unique relation between energy and momentum, the spectral weight sum rule is now valid for all states at a given momentum but not for all states at a given energy, and there is no reason to expect the same power law behavior as that at low energies.

In this paper, we compute the edge spectral function from a microscopic approach 
using the method of composite fermion (CF) diagonalization,
wherein we consider a truncated basis of states that contain no pairs of electrons with angular momentum unity. These are the only states that survive when the Haldane pseudopotential [\onlinecite{HaldanePseud, JainCFBook}] $V_1$ is taken to be infinitely strong; all states 
containing pairs with angular momenta equal to unity are pushed to infinity. 
Laughlin's 1/3 wave function [\onlinecite{Laughlin1983}] is exact for this model. Restriction to this subspace is also tantamount to considering edge excitations within the lowest $\Lambda$ level (or CF Landau level). A neglect of $\Lambda$ level mixing has been shown to be very accurate for the bulk physics, and we explicitly confirm below, by comparison to exact diagonalization results for small systems, that it provides a good first approximation for the edge excitations as well. Restricting to this truncated basis allows us to study systems with a large number of particles, providing better thermodynamic 
estimates than were available previously. 

We determine the thermodynamic limit of the dispersion of the elementary edge excitation. Our results show that while it is linear at low energies, it begins to deviate from linearity at an energy that is a fraction the 1/3 bulk gap. It also exhibits 
a roton minimum, which vanishes at a critical setback distance signaling edge reconstruction, in agreement with previous work 
[\onlinecite{Wan03,Jolad2}]. We show that while the spectral weights of individual excitations depend on various parameters [\onlinecite{Jolad2}], they 
accurately obey the sum rule mentioned above over the entire parameter range that we have studied. 

To determine the effect of the 
nonlinear dispersion on the edge exponent we employ a hybrid approach 
described in Sec.~V, wherein we build the spectrum from the elementary edge boson with a nonlinear dispersion but assume the spectral weights of the EFTA model. We 
evaluate the tunneling $I-V$ characteristic and 
find that the calculated exponent remains unchanged to a very good approximation even at energies above the edge roton energy where the dispersion is nonlinear. 
Z\"ulicke and MacDonald [\onlinecite{zuelicke96}] have also calculated the spectral 
function and the $I-V$ characteristics for a $\nu=1/3$ edge by assuming a 
dispersion $\epsilon(q) \sim -q 
\ln(\alpha q)$ for the edge magnetoplasmon, where $q$ is the momentum and $\alpha$ is a constant; they
have found that the edge exponent varies as the inverse filling. Our calculation 
is based on a magnetoplasmon dispersion that is obtained from a microscopic calculation for a system with Coulomb interaction and a realistic confinement potential. Recently, the effect of a nonlinearity of the
fermionic spectrum on the long-distance, low-energy correlation functions
has been studied in Refs. [\onlinecite{GlazmanScience09,GlazmanPRL09}].
However, this analysis considers a system in which both right and left-moving
modes are present and interact with each other, and it is not clear whether
the same analysis would be applicable to a FQH edge with a single chiral mode.

One may also expect some signature in tunnel transport that may be associated with the edge roton, which would then allow such transport to serve as a spectroscopic probe of the edge roton. However we find that the effect of edge roton on tunnel transport is negligible, because the spectral weight in the edge roton mode is very small.

Our paper is organized as follows. Section II contains a description of the model and the 
method of calculation. In Sec. III, we evaluate the energy spectra for small 
systems and compare them with the exact results. In Sec. IV, we 
study large systems and extract the thermodynamic edge dispersion, edge reconstruction and the edge roton. In Sec. V, we calculate the spectral weights and the associated sum rules for the EFTA, and we test the validity of these rules for the electronic spectra. We also outline our hybrid approach to calculate spectral function and tunneling density of states. In Sec. VI, 
we calculate the spectral function and tunneling density of states, present 
our main results on the $I-V$ characteristics, and mention their implications 
for the robustness of the edge exponent under a nonlinear dispersion. 
In Sec. VII, we discuss a system with a reconstructed edge 
using a field theoretic approach and address the effect on the tunneling 
exponent. We conclude in Sec. VIII with a summary and a discussion 
of the causes and implications of our main results. 
In the Appendix we give a mathematical formalism for the spectral weights, sum rules, and
the Green's functions for an ideal and a non-ideal EFTA.

\section{Model and Method of Calculation}

\subsection{Hamiltonian}

We consider a two-dimensional electron system in a plane.  The confinement is produced by a neutralizing background with uniformly distributed positive charge in a disk  (denoted $\Omega_N$)  of 
radius $R_N=(\sqrt{2N/\nu})\,l$; here  
$N$ is the number of electrons, $\nu$ is the filling factor, and $l=\sqrt{\hbar c/eB}$ is the magnetic length.  (The symbol $l$ is also used for angular momentum later, but the meaning ought to be clear from the context.) The background charge disk is separated  from the electron disk by a setback distance $d$. The ground state of 
the electron is determined by a microscopic calculation; we expect the 
electrons to be approximately confined to a disk of radius $R_N$ to ensure charge neutrality in 
the interior. This system is modeled by the following Hamiltonian: 
\bea H &\equiv& E_{\rm K}+V_{\rm ee}+V_{\rm eb}+V_{\rm bb} \non \\
&=& \sum_j\frac{1}{2m_b}\left(\vec{p}_j+\frac{e}{c}\vec{A}_j\right)^2 +
\sum_{j<k} \frac{e^2}{\epsilon |\vec{r}_j-\vec{r}_k|} \non \\
&& - \rho_0\sum_{j} \int_{\Omega_N} d^2r \frac{e^2}{\epsilon \sqrt{|\vec{r}_j
-\vec{r}|^2+d^2}} \non \\
&& + \rho_0^2 \int_{\Omega_N} \int_{\Omega_N} d^2rd^2r'\frac{e^2}{\epsilon 
|\vec{r}'-\vec{r}|}, \eea
where the terms on the right hand side represent the kinetic, 
electron-electron, electron-background, and background-background energies, 
respectively. Here $m_b$ is the band mass of the electrons, $\vec{p}_j$ is the
momentum operator of the $j$th electron and $\vec{r}_j$ is its position, 
$\vec{A}_j$ is the vector potential at $\vec{r}_j$, $\rho_0= \nu/2\pi l^2$ 
is the positive charge density spread over a disk of radius $R_N$, and 
$\epsilon$ is the dielectric constant of the background semiconductor material. At large 
magnetic fields, only the lowest Landau level states are occupied; hence the 
kinetic energy $\hbar \omega_c/2$ (where $\omega_c \equiv eB/m_b c$ is the 
cyclotron frequency) is a constant which will not be considered explicitly. 

\subsection{Electron States}

The single particle states in the $n^{th}$ Landau level are given, in 
the symmetric gauge, by
\beq \eta_{n,m}(z)=\frac{(-1)^n}{\sqrt{2\pi}}\sqrt{\frac{n!}{2^m(m+n)!}}
e^{-r^2/4}z^mL_n^m\left(\frac{r^2}{2}\right), \eeq
where $L_n^m(x)$ is the associated Laguerre polynomial 
[\onlinecite{Abramowitz}], $n$ and $m$ denote 
the Landau level index and angular momentum index respectively, $z=x-iy$ 
represents the electron coordinates in the complex plane, $r=|z|$, and all
lengths are quoted in units of the magnetic length $l$. The 
lowest Landau level states ($n=0$) are of special importance for our 
calculations below; they are given by
\beq \eta_{0,m}(z)=\frac{z^me^{-|z|^2/4}}{\sqrt{2\pi2^mm!}}. \label{wfLL} \eeq

The many-body states are formed by taking linear combinations of antisymmetric 
products of single particle wave functions denoted by $|p_1,p_2,\cdots p_N
\rangle=a_{p_1}^\dagger a_{p_2}^\dagger \cdots a_{p_N}^\dagger |0 \rangle$, 
where $p_i=\{n_i,m_i\}$ is the single particle state index of the $i^{th}$ 
electron, and $a_{p_i}^\dagger$ is the corresponding creation operator. 
We will be interested in the edge excitations of the FQH state at $\nu=1/3$ 
below. The ground state has total angular momentum $M_0=3N(N-1)/2$. The 
angular momentum of the excited state, $\Delta M$, will be measured 
relative to $M_0$. 

\subsection{Models of FQH Edge}

The tunneling of electrons from a two-dimensional electron gas into a 
Fermi liquid (such as a metal or $n+$ doped GaAs) has been studied 
experimentally in two geometries: point-contact geometry and cleaved-edge-overgrowth 
geometry [\onlinecite{Chang3}]. These are believed to represent realizations 
of smooth and sharp edges, respectively [\onlinecite{Grayson2}].

In the point-contact geometry, the boundary of the two-dimensional electron 
gas is smooth. Theoretically a smooth edge can be modeled by including all 
possible many-body edge excitation states for a given total angular momentum 
$M$ ($=\sum_{i=1}^{N} m_i$), placing no restrictions on the maximum single 
particle angular momentum $m_i$. The smoothness is ensured by states 
extending a few magnetic lengths beyond the disk edge. 

The cleaved-edge geometry is characterized by a long and thin tunneling 
barrier with a typical barrier width of about one magnetic length. Recent 
experiments suggest that the cleaved-edge-overgrowth represents the 
realization of a sharp quantum Hall edge [\onlinecite{Grayson2}]. A sharp 
edge can be modeled [\onlinecite{Wan03}] by excluding the single particle angular momenta 
beyond a cutoff $m_{max}$, given by 
\bea m_{max}=3(N-1)+l_0, \label{amcutoff} \eea
where $l_0$ is taken to be a small integer. 

We have calculated the edge spectra for both smooth edge and sharp edge, 
with cutoff $l_0=2$. We show in Sec. III, that the low-energy branch of the 
sharp edge matches with that of the smooth edge, and hence the edge dispersion
is not very sensitive to this issue. The calculations for the spectral functions are carried out for a smooth edge only. We note that a sharp edge eliminates several higher energy states, but does not significantly affect the low-energy branch and hence edge reconstruction. 

\subsection{Exact Diagonalization}

The exact interaction energy for FQH systems can be calculated for small 
systems using numerical diagonalization techniques.
In the disk geometry with symmetric gauge, for a given total 
angular momentum $M$, the basis states $| m_1,m_2,\cdots m_N\rangle$ in the 
lowest Landau level are generated according to the conditions:
\beq \sum_j m_j=M; \quad 0\le m_1< m_2 \cdots < m_N . \eeq
Restricting to the lowest Landau level, the Hamiltonian in the second 
quantized representation is
\bea H &=&\frac{1}{2}\sum_{r,s,t,u}\langle r,s|V_{\rm ee}|t,u\rangle a_r^\dagger 
a_s^\dagger a_t a_u \non \\
&& +\sum_m \langle m|V_{\rm eb}|m\rangle a_m^\dagger a_m+ V_{\rm bb}. \eea
Here the electron-electron and electron-background interaction matrix 
elements are defined as
\bea \langle r,s|V_{\rm ee}|t,u\rangle &=& \int d^2r_1 d^2 r_2\eta^*_{r}(r_1)
\eta^*_{s}(r_2)\frac{e^2}{\epsilon r_{12}}\eta_{t}(r_1)\eta_{u}(r_2); \non \\
\langle m|V_{\rm eb}|m\rangle &=& -\rho_0 \int d^2r_1\int_{\Omega_N} d^2 r_2 
\frac{|\eta_m(r_1)|^2}{\sqrt{r_{12}^2+d^2}}, \eea
with $r+s=t+u$. 

\begin{figure}[t] 
\begin{center}
\includegraphics[scale=1.0,viewport=60 620 402 790,clip]{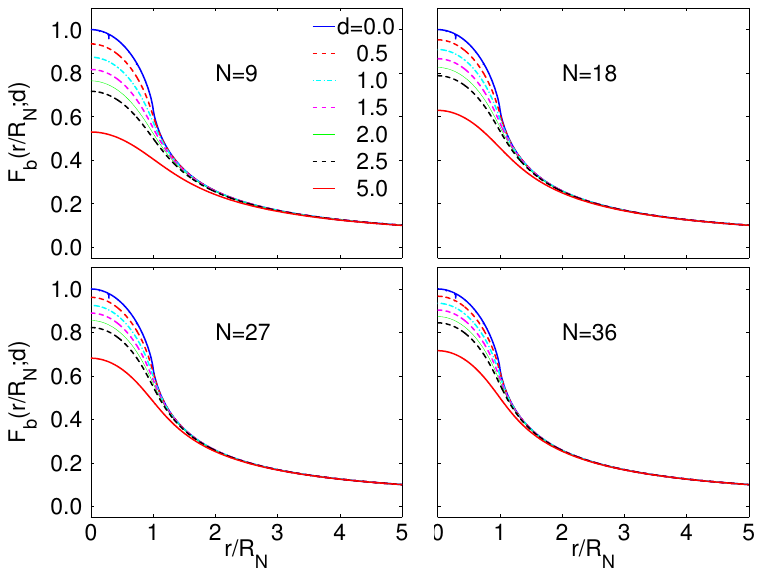}
\caption{(Color online) The function $F_b(r_i/R_N;d)$ that appears in the 
expression for the electron-background energy (Eq. (\ref{VebEq})) for several 
values of $d$ and $N$. Here $r$ is the distance of the electron from the 
center of the disk, and $R_N$ is the radius of the disk of the neutralizing 
positive background charge.} 
\label{Fig1}
\end{center} 
\end{figure}

The background-background interaction energy per particle is calculated 
analytically to be
\beq \frac{\langle V_{\rm bb} \rangle}{N}=\frac{\rho_0^2}{2N} 
\int_{\Omega_N}d^2r \int_{\Omega_N}d^2r' \frac{e^2}{\epsilon|\vec{r}-
\vec{r}'|} = \frac{8}{3\pi}\sqrt{\frac{\nu N}{2}}, \eeq
with the energy measured in units of $e^2/\epsilon l$. This adds a constant 
term to the matrix elements of the Hamiltonian. 
(The constant background-background interaction must be included to obtain 
a sensible thermodynamic limit for the energy, but is irrelevant for energy 
{\em differences}). Computing the electron-background interaction $\langle 
V_{\rm eb}\rangle$ requires numerical integration. To this end, we can write 
the electron-background energy as
\bea V_{\rm eb} &=& \sum_{i=1}^N v_{eb} (\vec{r}_i); \non \\
v_{eb}(\vec{r}_i) &=& -\rho_0 \int_{\Omega_N}d^2r
\frac{e^2}{\epsilon\sqrt{|\vec{r}_i- \vec{r}|^2+d^2}}\non \\ 
&\equiv& - \sqrt{2\nu N}F_b(r_i/R_N;d). \label{VebEq} \eea
For $d=0$, the integral in Eq. (\ref{VebEq}) on the right hand side can be 
calculated analytically, the result of which has been given by Ciftja and 
Wexler [\onlinecite{Ciftja}]. For $d \neq 0$, numerical integration is 
necessary. Figure \ref{Fig1} shows plots of the function $F_b$ for different 
$d$ and $N$.

\begin{center}
\begin{table*}[htb*]
\label{EDCFDDimTable}
\begin{tabular}{|c |@{\hspace{0.2in}} c @{\hspace{0.15in}} c @{\hspace{0.15in}}
c @{\hspace{0.15in}} c @{\hspace{0.15in}} c @{\hspace{0.15in}} 
c @{\hspace{0.15in}} c @{\hspace{0.2in}} | c @{\hspace{0.15in}} | }\hline 
$\Delta M$ & $D(N=6)$ & $D(N=7)$ & $D(N=8)$ & $D(N=9)$ & $D(N=10)$ & 
$D(N=11)$ &$D(N=12)$ & $D^{*}$ \\ \hline
0 &     1,206 &      8,033 &     55,974 &    403,016 &   2,977,866 &  
22,464,381 & 172,388,026 & 1 \\ 
1 &     1,360 &      8,946 &     61,575 &    439,100 &   3,218,412 &  
24,117,499 & 184,030,746 & 1 \\ 
2 &     1,540 &      9,953 &     67,696 &    478,025 &   3,476,314 &  
25,879,361 & 196,384,297 & 2 \\ 
3 &     1,729 &     11,044 &     74,280 &    519,880 &   3,752,096 &  
27,755,663 & 209,483,911 & 3 \\ 
4 &     1,945 &     12,241 &     81,457 &    564,945 &   4,047,402 &  
29,753,578 & 223,373,383 & 5 \\ 
5 &     2,172 &     13,534 &     89,162 &    613,331 &   4,362,833 &  
31,879,397 & 238,091,562 & 7 \\ 
6 &     2,432 &     14,950 &     97,539 &    665,355 &   4,700,201 &  
34,141,000 & 253,686,437 & 11 \\
7 &     2,702 &     16,475 &    106,522 &    721,125 &   5,060,174 &  
36,545,347 & 270,200,645 & 15 \\
8 &     3,009 &     18,138 &    116,263 &    780,997 &   5,444,732 &  
39,101,065 & 287,686,698 & 22 \\
\hline
\end{tabular}
\caption{Dimension $D$ of the full Hilbert space (used in exact 
diagonalization) for several values of $N$ and $\Delta M$. Here $\Delta M $ 
is the angular momentum measured relative to the angular momentum 
$M_0=3N(N-1)/2$ of the ground state of the $\nu=1/3$ FQH state. The 
last column gives $D^*$, the dimension of 
the CF basis in the lowest $\Lambda$ level, used in our CF diagonalization. 
The values of $D^*$ are given for sufficiently large $N$ where they are $N$ 
independent; for small $N$, $D^*$ may be smaller than the given value.}
\end{table*}
\end{center}

For the electron-electron interaction, we find the analytical expressions for 
$\langle r,s|V_{\rm ee}|t,u\rangle$ given by Tsiper in Ref. [\onlinecite{Tsiper}] to be 
useful. 
It is then straightforward to construct the Hamiltonian 
matrix and diagonalize it either by using
standard diagonalization procedures for small systems ($N\le 7$) to get the 
full spectrum, or by the Lanczos algorithm for slightly larger systems 
($N=8,9$) to get the low-energy spectrum. In the present work, we have 
performed full diagonalization for systems with up to 7 particles. 

\subsection{CF diagonalization}

We exploit the fact that the CF theory produces very accurate wave functions 
for low-energy eigenstates of the problem. Our approach will be to construct 
a truncated basis for the wave functions for the edge
excitations [\onlinecite{JainCFpaper,JainCFBook}], and then diagonalize the 
full Hamiltonian within this basis to obtain various quantities of interest.
The method has been described in detail in the literature 
[\onlinecite{Kamilla,MandalJain,GunSang}], so we present only a brief outline 
here. 

For the fraction $\nu=n/(2np+1)$, the CF theory maps interacting electrons at 
total angular momentum $M$ to non-interacting composite fermions at $M^*= M-
pN(N-1)$ [\onlinecite{Dev,Kawamura}] by attaching $2p$ flux quantum to each 
electron. The ansatz wave functions $\Psi^M_\alpha$ for interacting electrons 
with angular momentum $M$ are expressed in terms of the known wave functions 
of non-interacting electrons $\Phi^{M^*}_\alpha$ at $M^*$ as follows:
\beq \Psi^M_\alpha = {\cal P}_{\rm LLL} \prod_{j<k}(z_j-z_k)^{2p} 
\Phi^{M^*}_\alpha \label{PsiAlpha}. \eeq 
Here $\alpha=1, 2, \cdots, D^*$ labels the different states, 
${\cal P}_{\rm LLL}$ 
denotes projection into the lowest LL, and $D^*$ is the dimension of the 
CF basis. We choose $p=1$ as appropriate for $\nu=1/3$, and restrict 
$\Phi_\alpha^{M^*}$ to states with the lowest kinetic energy at $M^*$. No 
lowest Landau level projection is required for these states, as they are 
already in the lowest Landau level.

The Landau levels at $M^*$ transform into Landau-like effective kinetic energy
levels of composite fermions, called $\Lambda$ levels. The restriction to the
lowest Landau level at $M^*$ is equivalent to restricting composite fermions 
to their lowest $\Lambda$ level. More accurate spectra can be obtained by 
allowing $\Lambda$ level mixing and performing CF diagonalization (CFD) in a 
larger space, but that will not be pursued here. As will be seen below, the 
lowest $\Lambda$ level results are sufficiently accurate for our purposes.

The advantage of CF diagonalization is 
that the dimension $D^*$ of the CF basis is much smaller than the dimension of 
the full lowest Landau level Hilbert space at $M$; this allows a study of 
much larger systems. Table I compares the dimensions of the full Hilbert 
space ($D$) and the truncated CF space ($D^*$) for 6 to 12 particles 
for several values of $\Delta M$. The dimension $D$ increases exponentially, 
approximately as $D=10^{-2}\exp(2N)$ for large $N$. This gives $D\approx 
2\times 10^{29}$ for $N=36$ particles, in dramatic contrast to $D^*\sim$ 
10-100 for $0\le \Delta M<10$. Of course, the Hilbert space reduction 
comes with a cost: the CF basis functions are much more complicated than the 
usual single Slater determinant basis functions, and diagonalization of the 
Hamiltonian in this basis requires many non-trivial steps and extensive Monte 
Carlo. Nonetheless, CF diagonalization can be, and has been, performed for 
many non-trivial cases of interest.

We need to evaluate the matrix elements of the Hamiltonian in our CF basis. 
If $\Psi_\alpha^M(z_1,z_2,\cdots,z_N)$ and $\Psi_\beta^M(z_1,z_2,\cdots,z_N)$ 
denote two CF states at angular momentum $M$, then the electron-background 
and electron-electron energy matrix elements are given by $\langle \Psi_\alpha^M |V_{\rm eb} | \Psi_\beta^M \rangle$ and $\langle \Psi_\alpha^M |V_{\rm ee} |\Psi_\beta^M \rangle$. 
Their evaluation requires evaluating 
multi-dimensional integrals, which can be effectively accomplished by 
Monte Carlo techniques described in the next section. The CF basis functions 
are in general not orthogonal to each other. They can be orthogonalized by 
the Gram-Schmidt procedure adapted for CF states to produce the energy 
spectrum as described in the literature 
[\onlinecite{Kamilla,MandalJain,GunSang, JainCFBook}]. Essentially, given 
the interaction matrix 
${\hat V}_{\alpha, \beta}=\langle \Psi_\alpha |V | \Psi_\beta \rangle$ and the
overlap matrix ${\cal{O}}_{\alpha, \beta}=\langle \Psi_\alpha | \Psi_\beta 
\rangle$, the energies and eigenvalues are obtained by diagonalizing 
the matrix ${\cal{O}}^{-1} {\hat V}$ .

\subsection{Monte Carlo Methods}

Multi-dimensional integrals
can be evaluated most effectively by the Metropolis-Hastings Monte Carlo (MHMC) 
algorithm [\onlinecite{metropolis,Hastings, ChibMetropolis}]. For a 
discussion of the application of MHMC algorithm to quantum many-body systems, 
in particular to quantum Hall systems, we refer the reader to Refs. 
[\onlinecite{JainCFBook,Ciftja}]. 
For our energy calculations, we find it sufficient to thermalize
for 100,000 iterations and then average over about $10-20$ million iterations 
for each angular momentum. For spectral weights calculations, about 200 
million iterations are required for the eigenvector. These numbers do not 
vary significantly with $N$ in the range of our study ($N\leq 45$), but the 
computation time increases exponentially 
with $N$ and $\Delta M$, limiting our study to systems up to $N=45$ particles 
for the energy spectrum, and $N=27$ for spectral weights. The energies were 
calculated for $\Delta M=1-8$ and the spectral weights for $\Delta M = 1-4$.
 
\section{Small system studies}

\begin{figure}[t] 
\begin{center}
\includegraphics[scale=1.0,viewport=60 625 402 790,clip]{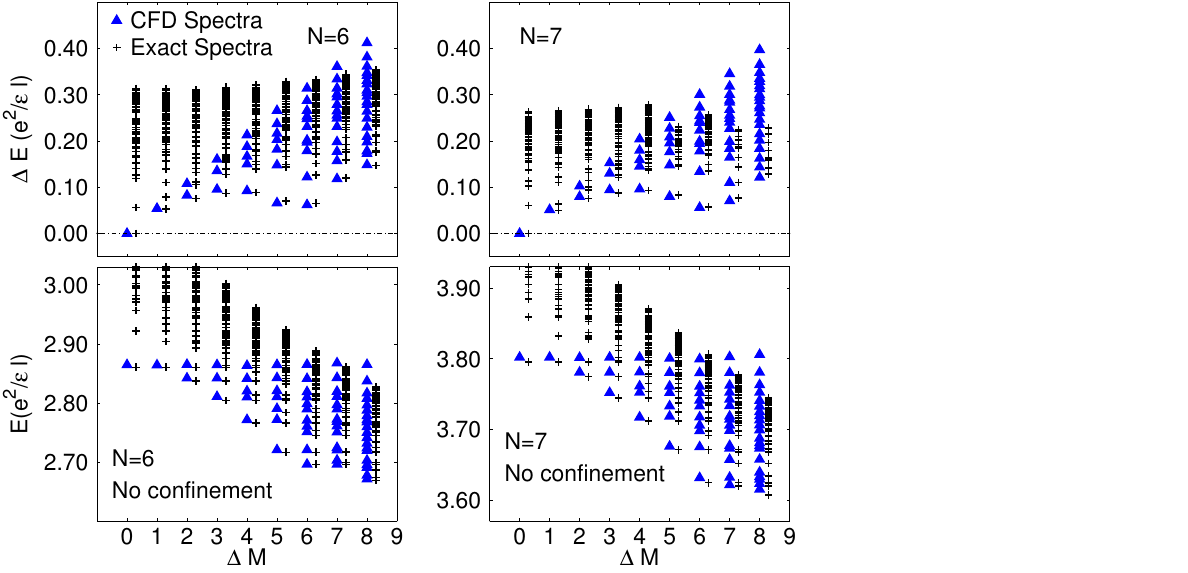}
\caption{(Color online) Comparison of CFD spectra with the exact spectra for 
the smooth edge for 6 and 7 particles, with and without a positive background 
(upper panels and lower panels, respectively). For the upper panels, we have 
taken $d=0$. Blue triangles are the energies obtained by CF diagonalization,
and the adjacent `+' symbols (shifted along the $x$ axis for clarity) are the 
exact energies. The high energy parts of the exact spectrum are not shown. 
All energies are quoted in units of $e^2/\epsilon l$, and are measured 
relative to the energy of the ground state at
$\Delta M=0$. $\Delta M$ is the angular momentum of the excitation.}
\label{Fig2} 
\end{center} 
\end{figure} 

\begin{figure}[htbp] 
\begin{center}
\includegraphics[scale=1.0,viewport=60 625 402 790,clip]{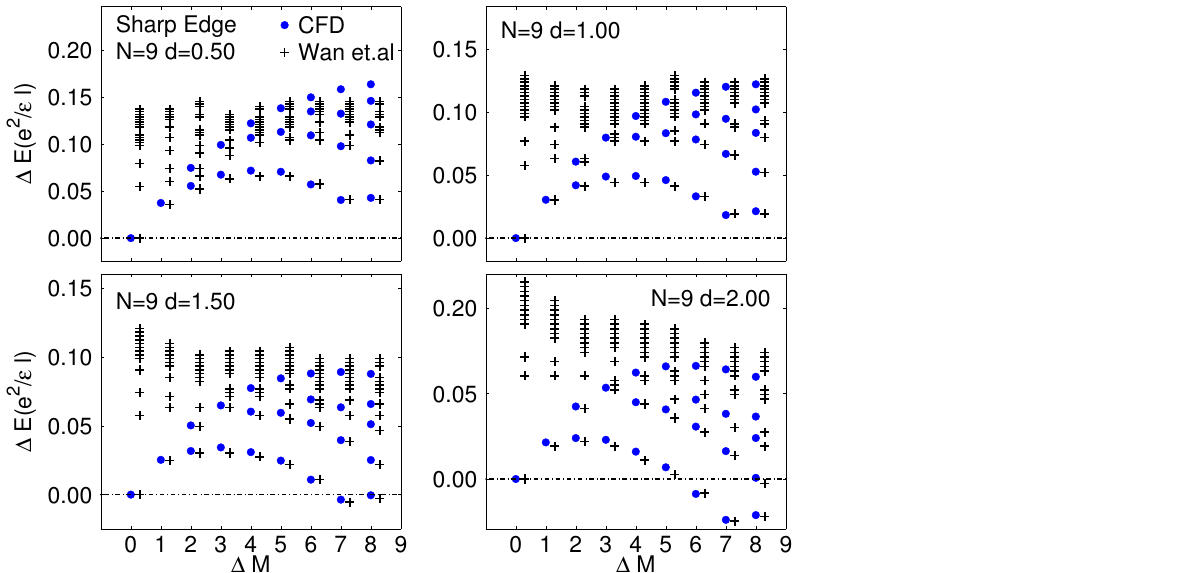}
\caption{(Color online) Comparison of the CFD spectra with the exact spectra 
for a sharp edge for $N=9$ particles at $\nu=1/3$ with $d=0.5-2.0$ $l$. 
Blue dots indicate the energies obtained by CF diagonalization, whereas the 
adjacent `+' symbols (shifted along the $x$ axis for clarity) are energies 
extracted from the figures of Ref. [\onlinecite{Wan03}].} 
\label{Fig3} 
\end{center} 
\end{figure}

The CFD approach has been well tested in the past for the {\em bulk} physics 
at various filling factors and has been shown to capture the behavior of 
FQH systems accurately. Before proceeding to larger systems, we first test 
the validity of the CFD approach for the edge excitations. 

Using the CF diagonalization procedure outlined earlier, we compute the edge 
excitation spectra for the $\nu=1/3$ state for several parameters in the range
$d=0$-$2.5l$ and $\Delta M =0$-$8$. Figures \ref{Fig2} and \ref{Fig3} show 
comparisons of the CFD spectra with the exact spectra for smooth and sharp 
edges (exact diagonalization is possible for slightly larger systems for a 
sharp edge because of the additional restriction on the Fock space),
demonstrating that the CFD approach is essentially exact. For a sharp edge, 
we have chosen the value $l_0=2$ (cf. Eq. (\ref{amcutoff})). The exact 
diagonalization results in Fig. \ref{Fig3} are taken from Wan {\em et. al} 
[\onlinecite{Wan03}]. Consistent with their conclusions, we find that 
edge reconstruction occurs for $d$ greater than a critical separation.

We note that for small systems the results for sharp and smooth edges
are not very different, as shown in Fig. \ref{Fig4}; both show edge 
reconstruction for $d>1.5l$. For larger systems, as seen below, edge 
reconstruction occurs for larger $d$ for the sharp edge as expected.

\begin{figure}[t]
\begin{center}
\includegraphics[scale=1.0,viewport=60 545 402 790,clip]{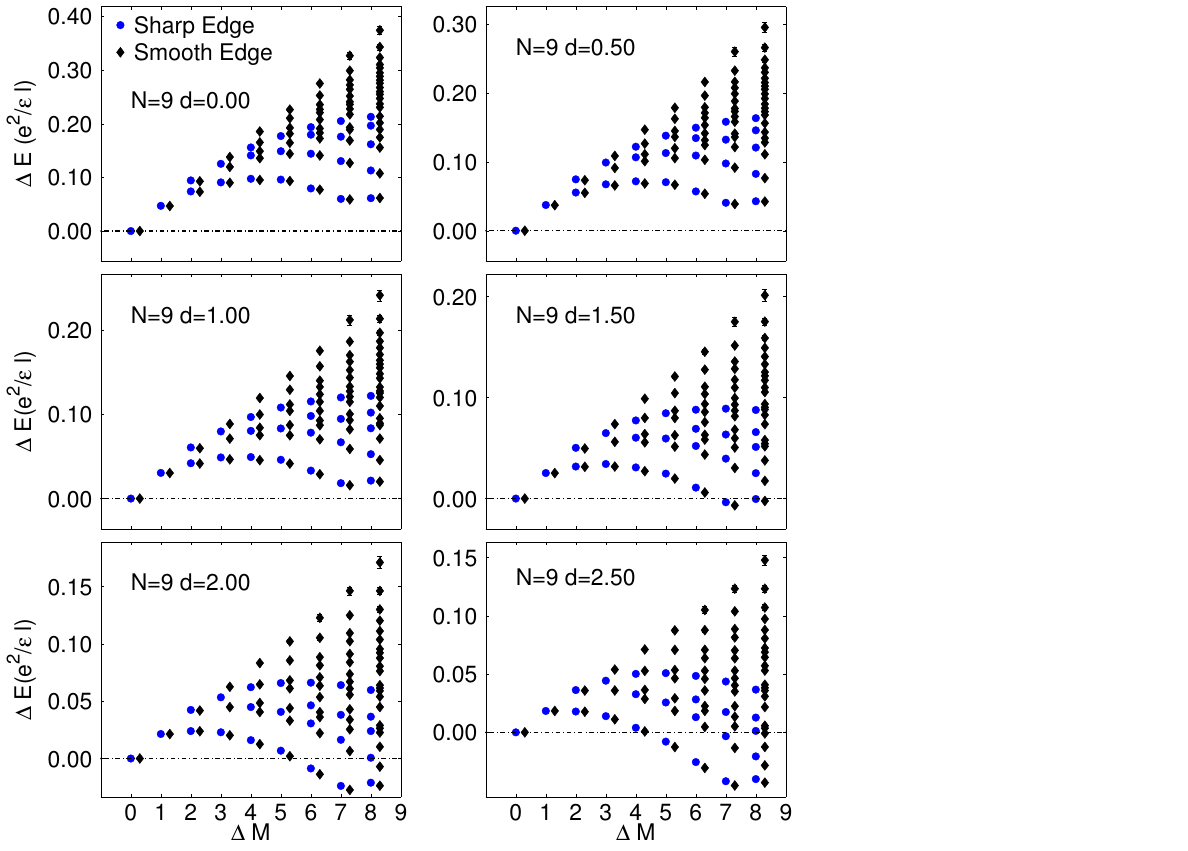}
\caption{(Color online) Energy spectra for smooth and sharp edge excitations 
of a $\nu=1/3$ system with $N=9$ and $d=0.0-2.5$ $l$. Blue dots are for the 
sharp edge, whereas the adjacent black diamonds (shifted along the $x$ axis 
for clarity) are for the smooth edge.}
\label{Fig4}
\end{center}
\end{figure}

\section{Spectra and Edge dispersion}

Having ascertained the validity of our approach from comparisons to exact results, we now proceed to investigate the physics in the thermodynamic limit. We 
study the edge spectra for different sizes and approach the thermodynamic limit
by identifying a scaling relation between the physical momentum $\delta k$ 
(we take $\hbar=1$) and the edge angular momentum $\Delta M$. The momentum is 
related to the size of the system by $k\sim r/l^2$, where $r$ is the 
radius of the orbital wave function. For edge electrons, $r=\sqrt{2M} 
l=\sqrt{2(3(N-1)+\Delta M)}l$ for a system with $\nu=1/3$. This gives 
the momentum of the edge excitation to be 
\beq \delta k= \frac{\Delta M}{\sqrt{6(N-1)} l}. \label{ScaleMomenta} 
\eeq
Henceforth we will denote $\delta k$, the physical momentum of the edge 
excitation, as simply $k$.

\begin{figure*}[htbp]
\includegraphics[scale=1.0,viewport=60 480 575 790,clip]{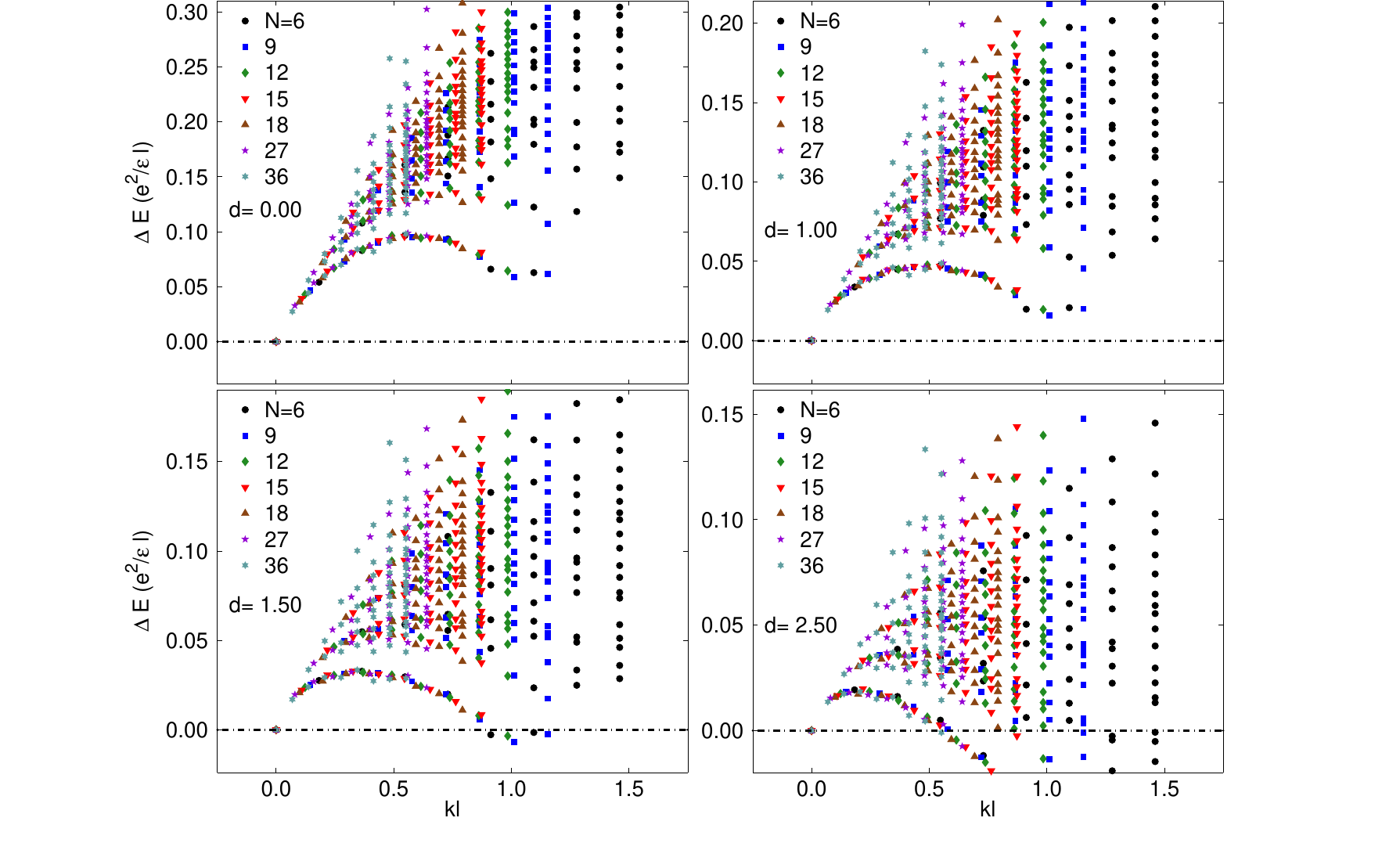}
\caption{(Color online) Edge spectra as a function of the physical momentum 
(see Eq. (\ref{ScaleMomenta})) for $N=6-36$ particles. The setback distance $d=0.0-2.5 l$. 
Data collapse for the lowest spectral branch can be seen in all the panels. Lower panels with $d\ge1.5l$ show edge reconstruction.}
\label{ThermSpectraPlots}
\end{figure*}

Based on our edge spectra results in Fig. \ref{ThermSpectraPlots}, with the 
parameters $N=6-36$, $d=0$-$2.5l$, and $\Delta M =0$-$8$, we make the 
following observations.

\begin{figure*}[t]
\includegraphics[scale=1.0,viewport=60 480 575 790,clip]{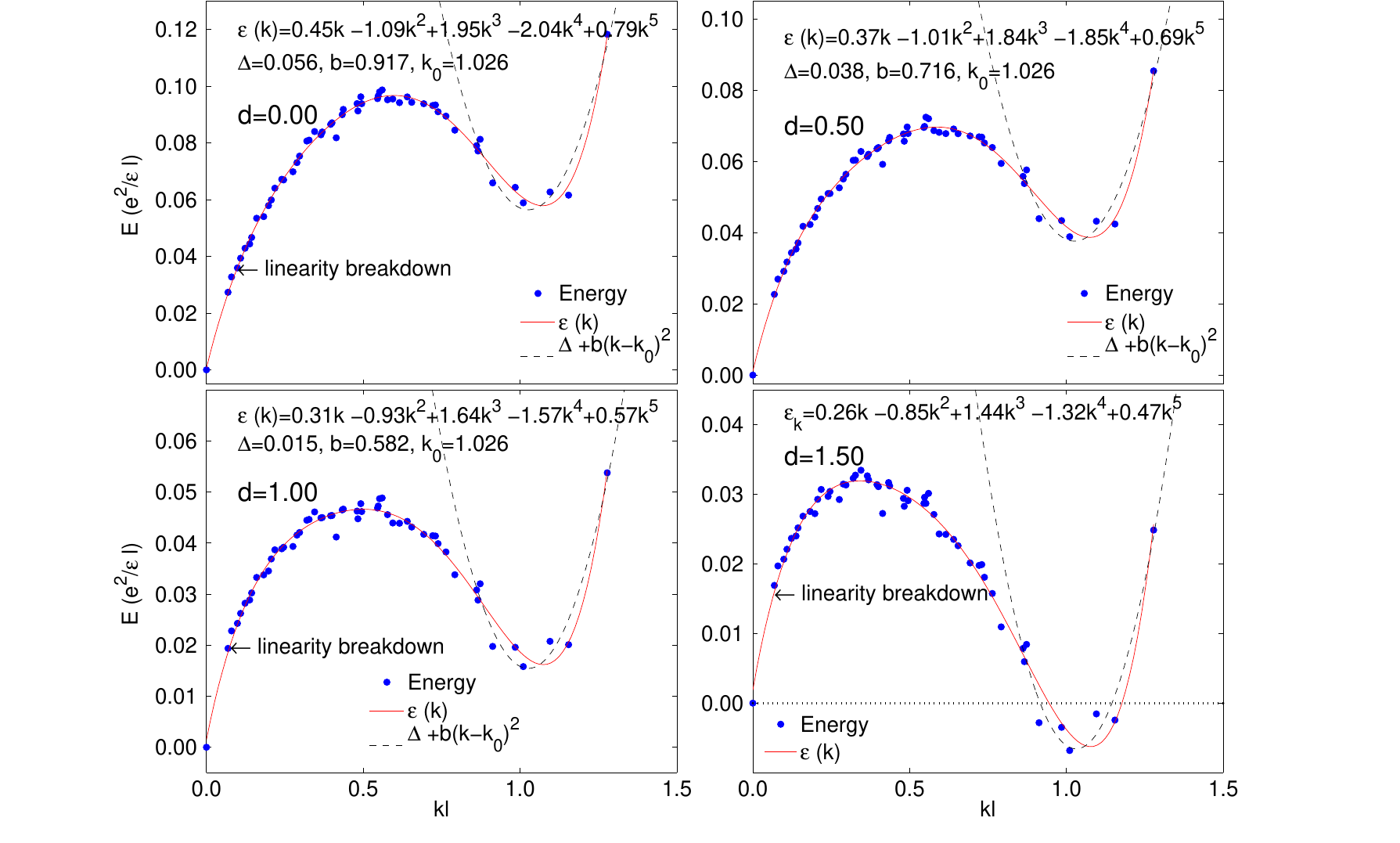}
\caption{(Color online) Dispersion $\varepsilon (k)$ of single edge boson for different setback distances in the range $d=0-1.5l$. The solid lines are a fifth order polynomial fit to the lowest branch of the energy spectra in Fig. \ref{ThermSpectraPlots}. Arrows indicate the energy beyond which the dispersion becomes nonlinear, 
as defined in section IV. Dispersion minima for the un-reconstructed edges ($d<1.5l$) have been fitted with a roton curve $\varepsilon_{roton} (k)=\Delta +b(k-k_0)^2$. Roton gap, momenta and curvature corresponding to the roton minima are also shown. Note that Fig. \ref{ThermSpectraPlots} has $d$ up to $2.5l$, but in this figure we show the plots for the relevant distance range $d=0-1.5l$.} 
\label{DispersionFits}
\end{figure*}

\begin{enumerate}

\item {\bf Data Collapse}: The energy spectra for different system sizes 
collapse, indicating proper scaling to the thermodynamic limit. The lowest 
branch in each of the four panels corresponds to the dispersion of the single 
edge boson, for various 
setback distances in the range $d=0-2.5 l$. The data collapse to a single curve is apparent
even for the second lowest branch, beyond which energies form a continuum. 
A few points for $N=36$ deviate slightly from the common trend in the lowest 
branch, which we believe is due to convergence problems for larger systems. 

\item {\bf Edge Reconstruction}: For $d>d_c$, we observe edge reconstruction 
due to competing electron-background energy and electron-electron interaction 
energy. 

\item {\bf Nonlinearity and Edge Rotons}: The lowest branch, though 
linear at low $k$, eventually deviates from linearity for all $d$. We extract 
in detail the dispersion of the single boson excitation for various $d$ values in 
Fig. \ref{DispersionFits}, with polynomial fits shown on the plots themselves. We observe that the edge dispersion is nonlinear and the ``linearity 
breakdown" (defined as the point at which the deviation is $\sim$20\% from linear) occurs at energies in the range of
$0.02-0.04 (e^2/\epsilon l)$; in experiments, this corresponds to the 
range $0.2$meV to $0.4$meV. For $d<d_c\approx 1.5l$, the dispersion also shows a roton structure with the minima around $k_0=1.026l^{-1}$. The roton gap
$\Delta_R$ is approximately $0.056 (e^2/\epsilon l)$ for zero setback distance, but 
depends on the setback distance and collapses at approximately $d_c=1.5l$.
The analytical fits for the dispersion relations are given in Fig. \ref{ThermSpectraPlots}.

\end{enumerate}

The nonlinear dispersion and the existence of the edge roton lie outside the assumptions 
of the EFTA model. In the next two sections we explore their effect on 
the edge exponent that is relevant to tunneling into the edge.

\section{Bosonization of FQH edge}

The bosonic EFTA model is based on the idea that the edge excitations can be 
mapped into excitations of a bosonic system, given by 
\beq |\{ n_l \}\rangle = \prod_{l=0}^{\infty} \frac{b_l^{n_l}}{\sqrt{n_l!}}|0
\rangle \;, 
\label{BosState} \eeq
where $n_l$ is the number of bosons in the orbital with angular momentum $l$.
For a given state $\{ n_l \}$, the total angular momentum and total energy are
given by
\bea \Delta M &=& \sum_l l \,n_l, \non \\
E_{\{n_l\}} &=&\sum_l n_l \epsilon_l. \label{BosEnergies} \eea
Furthermore, the electron field operator at filling factor $\nu=1/m$ is 
given by [\onlinecite{WenIntJModPhy}]
\beq \hat{\psi}^\dagger (\theta) \propto e^{-i\sqrt{m}\hat\phi(x)}= 
\sqrt{\eta} e^{-i\sqrt{m}\hat{\phi}_+(\theta)} e^{-i\sqrt{m}
\hat{\phi}_-(\theta)}, \label{FieldOp} \eeq
where $\sqrt{\eta}$ is a normalization factor. The fields 
$\hat{\phi}_+(\theta)$ and $\hat{\phi}_-(\theta)$ can be expanded in terms of 
bosonic creation and annihilation operators $b_l$ and $b_l^\dagger$ as
\bea \hat{\phi}_+(\theta)&=-&\sum_{l>0} \frac{1}{\sqrt{l}} b_l^\dagger 
e^{il\theta} \non \\
\hat{\phi}_-(\theta)&=-&\sum_{l>0} \frac{1}{\sqrt{l}}b_l e^{-il\theta}.
\label{FieldExp} \eea 

\subsection{Electronic and Bosonic edge spectra}
\label{sec:BosSpectrum}

We first ask if the excitation spectrum of the electronic problem conforms to 
the bosonic prediction, in which all excitations are created from a single 
branch of bosons. Following Ref. [\onlinecite{Wan03}], we identify the lowest 
energy state at each angular momentum $\Delta M$ in the electronic spectrum 
with a single boson excitation at $l=\Delta M$, i.e., $n_l=
\delta_{l,\Delta M}$. This gives the energy dispersion $\epsilon_l$ of the 
single boson state as a function of $l$, where we measure the energy 
$\epsilon_l$ with respect to the energy at $\Delta M=0$ ($M=M_0$). Using 
the equations $\sum_l l n_l =\Delta M$ and $E_{\{n_l\}}=\sum_l n_l \epsilon_l$,
the energies of all the bosonic states $\{n_l\}$ can be obtained and 
identified with the energies of the corresponding electronic states. We note 
that in our truncated basis, the numbers of CF and bosonic states are equal 
at each $\Delta M$. 

\begin{figure}[htbp]
\begin{center}
\includegraphics[scale=0.46,viewport=55 580 570 790,clip]{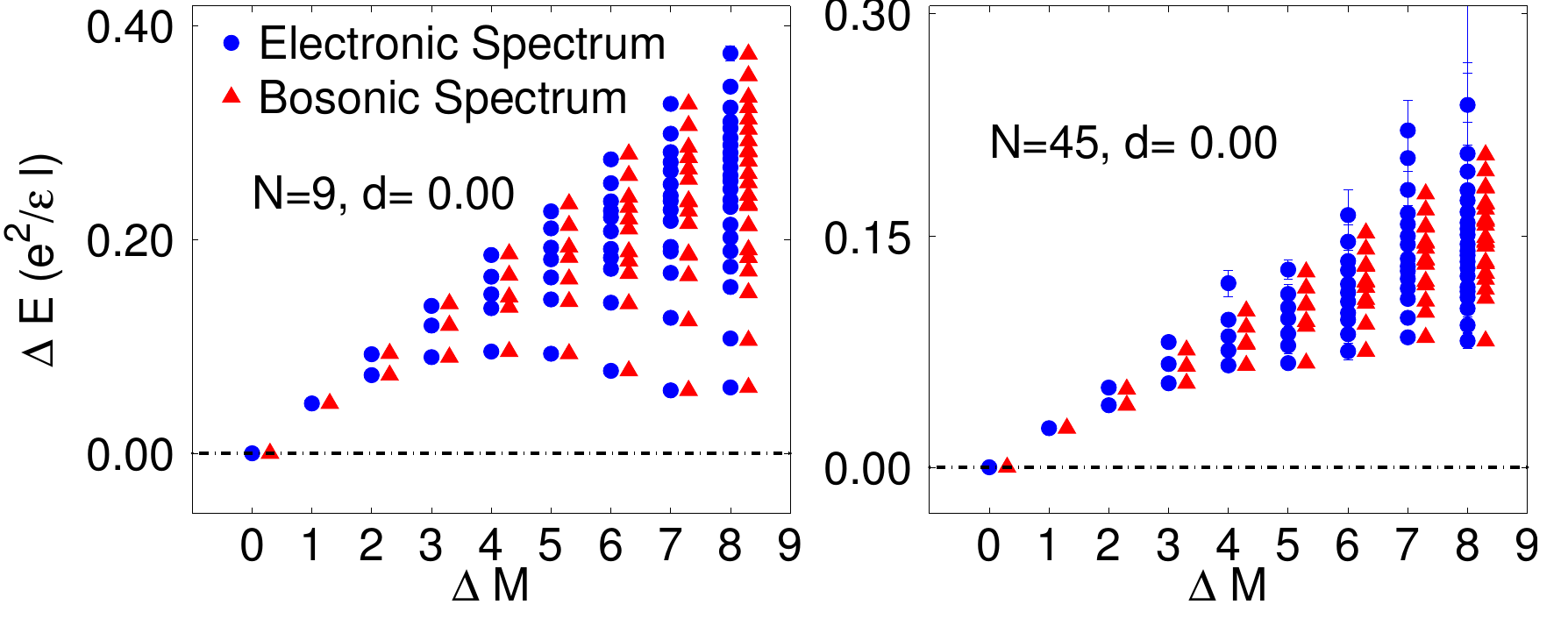}
\caption{(Color online) Energy spectrum for the edge excitations of $\nu=1/3$ 
(blue dots), obtained by CF diagonalization, for $N=9,45$ at $d=0.0$. Red 
triangles (shifted along the $x$ axis for clarity) show the bosonic spectra 
generated from its lowest branch (see section V.A for explanation).}
\label{Fig7}
\end{center}
\end{figure}

In Fig. \ref{Fig7}, we compare the bosonic excitation spectrum obtained in 
this manner with the electronic spectra computed through CFD for the edges 
for the cases $N=9,45$ and $d=0.0$. The CFD spectra are shown in blue circles 
and the bosonic spectra are shown in red triangles. In all cases, the spectra 
obtained from the bosonic picture, with the single boson dispersion as an 
input, show a close resemblance to the electronic spectra, confirming the 
bosonic picture as well as the interpretation of the lowest branch as the 
single boson branch. (The bosonic description becomes less accurate with 
increasing $N$ or $\Delta M$, but still remains accurate for the low-energy 
states).

\subsection{Spectral Weights}
\label{sec:SpecWt}

The relation between the electron and the boson operators given in Wen's 
ansatz in Eq. (\ref{Wenansatz}) leads to a precise prediction for the 
matrix elements of the electron field operator. We will study, following 
Palacios and MacDonald [Ref. \onlinecite{PalaciosMacDonald}], these matrix 
elements, called spectral weights, defined by 
\beq C_{\{n_l\}}= \frac{\langle \{n_l\} | \hat\psi^\dagger(\theta)|0\rangle}
{\langle 0 | \hat\psi^\dagger(\theta)|0\rangle}, \label{spectralweight} \eeq
where $|\{n_l\}\ra$ represents the bosonic state with occupation $\{n_l\}$, 
$|0\ra$ is the vacuum state with zero bosons, $\hat\psi^\dagger(\theta)$ 
is the electron creation operator at position $\theta$ (with one dimension 
wrapped into a circle), and $l$ denotes the single boson angular momentum.

Using Eqs. (\ref{Wenansatz}), (\ref{FieldExp}) and (\ref{BosState}), 
it is straightforward to obtain the EFTA predictions for the spectral weights 
\beq |C_{\{n_l\}}|^2=\frac{m^{n_1+n_2+\cdots}}{n_1!n_2!\cdots 1^{n_1} 2^{n_2}
\cdots}. \label{SqSpectWt} \eeq
We note that the denominator in Eq. (\ref{spectralweight}) eliminates the 
unknown normalization constant $\sqrt{\eta}$ in Eq. (\ref{FieldOp}).

To obtain the spectral weights from our electronic spectra, we need to
identify a ``dictionary" between the bosonic states and the electronic states.
It is natural to identify the vacuum state $|0\rangle$ with the ground state of
interacting electrons at $\nu=1/m$, denoted by $|\Psi_0^N \rangle$. The 
field operator has the standard meaning of
\beq \hat\psi^\dagger(\theta)=\sum_l\eta_l^*(\theta) a_l^\dagger \equiv \sum_l
\psi_l^\dagger(\theta), \eeq
where $a_l^\dagger$ and $a_l$ are the creation and annihilation operators for 
an electron in the angular momentum $l$ state, the wave function for which is 
given in Eq. (\ref{wfLL}). The wave function $\Psi_{\{n_l\}}^{N+1}(\{z_i\})$ 
is the electronic counterpart of the bosonic state $\vert {\{n_l\}} \rangle$ 
obtained through CF diagonalization. Using these definitions we calculate the electronic spectral weights. 
The details of the mapping and calculational method have been discussed in a previously published work [\onlinecite{Jolad2}]. 

\begin{figure*}[tp]
\includegraphics[scale=1.0,viewport=60 665 575 790,clip]{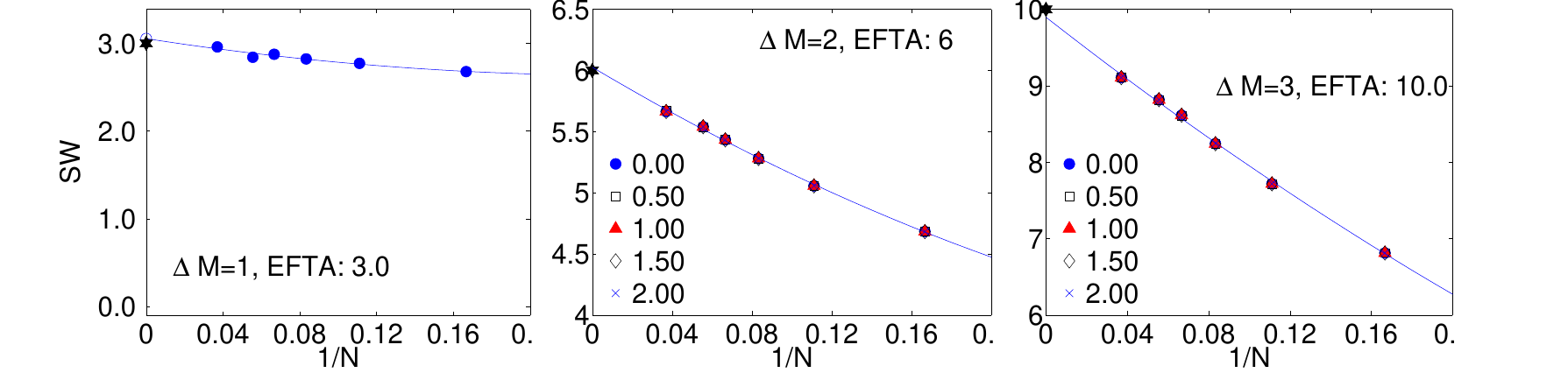}
\caption{(Color online) Sum of spectral weights (Eq. (\ref{SqSpectWt})) obtained from the electronic spectra (see section V.B for definition) at angular momenta $\Delta M=1-3$ and $d=0-2.0l$. For $\Delta M=1$, there is only one one state, which is independent of $d$ in our model. For other cases we find that the spectral weight sum is independent of $d$. Also, the thermodynamic limit is consistent with the sum rule derived from the EFTA (see Eq. (\ref{SSWGenl})).}
\label{SWsumPlots}
\end{figure*} 

\subsection{Spectral Weight Sum rules}
\label{sec:SpecWtSumrule}

As seen below, a sum rule for the spectral weights plays an important role. 
For $\nu=1/m$, in the bosonic EFTA, the sum of the squared spectral weights (SSW)
is given by (see Appendix A for a derivation), 
\bea SSW_{\Delta M}^{EFTA} 
&=& \frac{(\Delta M+m-1)!}{\Delta M!(m-1)!}, \non \\
\sum_l l n_l &=& \Delta M. \label{SSWGenl} \eea
It is natural to ask whether the above relation holds 
for the real FQH edge. We test the validity of the sum rules for 
$\nu=1/3$ in our model of a FQH edge by computing the spectral weights for 
system sizes $N=9-27$ and $\Delta M=1-3$. The results for individual 
spectral weights have been published in a previous work by two of the authors
[\onlinecite{Jolad2}]. In Fig. \ref{SWsumPlots}, we show the plots of the 
SSW for Coulomb interactions for different $N$. The thermodynamic limit for 
the SSW approaches the expected result according to Eq. (\ref{SSWGenl}). 

\subsection{A Hybrid Model}
\label{hybrid}

To obtain results for the
spectral function and the tunneling density of states in the parameter regime of our interest, bigger systems and larger angular momenta are needed. 
We have found that it is computationally infeasible to 
calculate the spectra for $N\gtrsim50$ and $\Delta M \gtrsim 8$, and
spectral weights for $N\gtrsim 27$ and
$\Delta M \gtrsim 4$. To make further progress we used a hybrid
approach. We work with the single boson dispersion obtained by the microscopic 
theory, but we assume that (i) the full spectrum can be constructed from it by assuming 
that the bosons are noninteracting, and (ii) the spectral weights of individual 
states are given by the EFTA model. With these assumptions, our model tests 
only the effect of nonlinearity of the single boson dispersion. Corrections to the edge 
exponent arising from coupling to states outside of our restricted basis, as well as those from a redistribution of the spectral weights between states, are outside the scope of our present study.

As an illustration of our hybrid approach, we have plotted in panel (c) of Fig. 
\ref{SFFigN36} the spectral weights of various excited states 
discussed in Sec. \ref{sec:BosSpectrum}. The figure illustrates 
that the spectral weights corresponding to a given number of 
bosons have roughly the same energy, in agreement with previous work by 
Z\"ulicke and MacDonald [\onlinecite{zuelicke96}]. 

\section{Spectral Function and Tunneling Density of states}

The positive energy part of the electron spectral function is given by 
[\onlinecite{GiulianiVignale,JainCFBook}], 
\beq A^>(k,E)=\sum_\alpha \vert \langle \alpha, N+1\vert c_k^\dagger \vert 
0,N\rangle \vert^2\delta (E-E_\alpha^{N+1}+E_0^N), \label{SpecFun} \eeq
where $\alpha$ denotes many-body energy eigenstates, and $k,c_k^\dagger$ 
denote the momentum (or any other) quantum number and the corresponding 
electron 
creation operator respectively. For the FQH edge, if we restrict to the 
states in the lowest Landau level, $\Psi_{\{n_l\}}^{N+1}(\{z_i\})$ would 
correspond to $\vert \alpha, N+1 \rangle$. Using the definition of the 
spectral weight, we can write the spectral function as 
\bea A^>(k,\epsilon)&=& \eta \sum_{\{n_l\}} \vert C_{\{n_l\}} \vert^2\delta 
(\epsilon- E_{\{n_l\}}^{N+1}), \non \\ 
\sum_l l n_l &=& \lambda^{-1}k, \label{SpecFunedge} \eea
where $\vert C_{\{n_l\}} \vert^2$ is the electronic spectral weight,
$ E_{\{n_l\}}^{N+1}$ is the energy of the electronic spectra measured 
from the ground state of $N+1$ particles, and $\epsilon$ is the energy of the 
edge excitation measured with respect to the chemical potential $\mu$. Here, 
$\eta$ is the normalization factor $\vert \langle 0 | \hat\psi^\dagger(\theta)
|0 \rangle \vert^2$ in Eq. (\ref{spectralweight}). We divide the energy into discrete bins $[\epsilon-\delta/2,\epsilon+\delta/2)$ of width $\delta$ and sum over the spectral weights for states with the corresponding energies and momentum $k$ to calculate $A(k,\epsilon)$. As discussed in Sec.\ref{hybrid}, we have used the electronic energy dispersion and the bosonic spectral weights to calculate the spectral function.

\begin{figure*}[t]
\centering
\includegraphics[scale=1.0,viewport=60 450 575 790,clip]{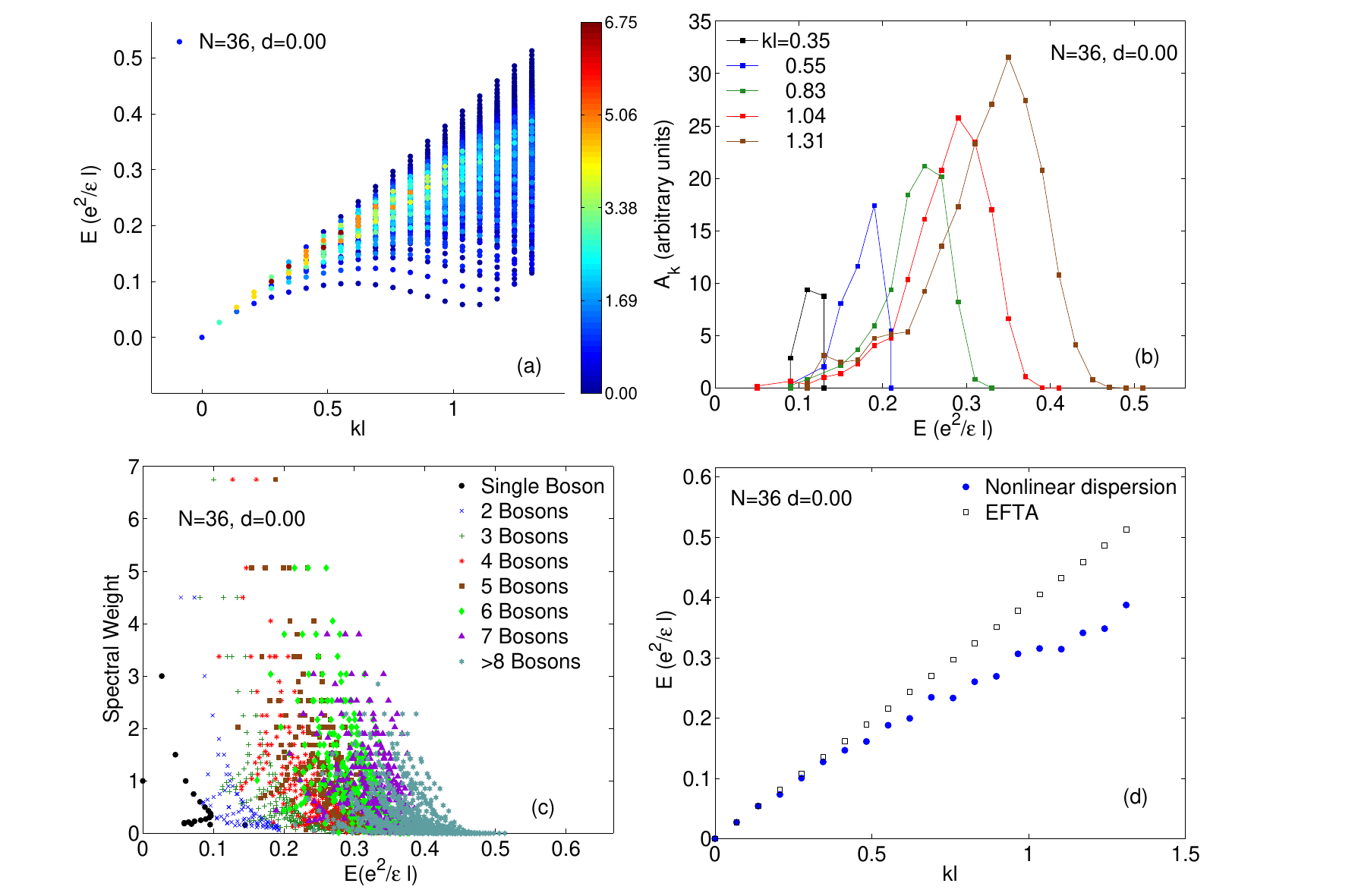}
\caption{(Color online) (a) Bosonic spectra generated through electronic 
dispersion for $N=36$ particles and $d=0.0$ using the hybrid approach described in Sec. V.D; the bosonic spectral weights are shown in graduated colors. (b) Spectral function calculated from Eq. (\ref{SpecFunedge}) for different $k$ using the data in panel (a). (c) Bosonic spectral weights plotted as a function of the energy, grouped by the number of bosons. (d) Energy at the maxima of the spectral function for different $k$. The black empty squares indicate the situation for a linear dispersion corresponding to EFTA. The system has $N=36$ particles; the setback distance is $d=0.0$; and we have restricted to angular momentum up to $\Delta M_{max}=19$.}
\label{SFFigN36}
\end{figure*}

In Fig. \ref{SFFigN36}, panel (a), we show the energy spectra with spectral 
weights (colored) for bosonic states. The low-energy states have 
comparatively smaller weight. The spectral function $A(k,\epsilon)$ (unnormalized
and in arbitrary units) for different momenta $k$ are shown in panel (b). In panel (d) we note that the energy corresponding to the maximum of spectral function closely follows the line of maximum energy for a given momenta.

\subsection{Tunneling density of states}

When an electron tunnels between two weakly coupled systems (labeled 
$L,R$) with a chemical potential difference $eV$, the tunneling current can 
be shown to be ([\onlinecite{GiulianiVignale,JainCFBook}]), 
\beq I(eV) \sim \sum_{\alpha, \beta}\vert T_{\alpha, \beta} \vert^2 
\int_0^{eV}dEA_L(\alpha,E)A_R(\beta,eV-E), \eeq
where $\alpha, \beta$ are the quantum numbers of the electron states
in the two systems, and $T_{\alpha, \beta}$ is the matrix element 
connecting the two states. If the energy range of tunneling is small, 
$T_{\alpha, \beta}$ can be approximated by a constant $T$ independent of 
the quantum numbers. Further assuming one 
system (say $L$) is a metal, whose the density of states is 
almost constant near the Fermi surface, gives 
the differential conductance as proportional to the tunneling 
density of states in the other system ($R$, labeled as ``edge") as 
\beq \frac{dI}{dV}\Big |_{\rm metal- edge}\sim D_{\rm edge}(eV)\equiv \sum_\alpha A(\alpha, E). \eeq

For a FQH edge, using Eq. (\ref{SpecFunedge}), the tunneling density of states 
(the superscript $N+1$ is omitted for brevity) is given by,
\beq D_{\rm edge}(\epsilon)\sim \sum_{\{n_l\}} \vert C_{\{n_l\}} \vert^2\delta 
(\epsilon- E_{\{n_l\}}). \eeq
The relation between $I$ and $V$ is given by 
\bea I(eV)&\sim& \sum_k \int_0^{eV} d\epsilon A^>(k,\epsilon) \non \\
&\sim& \int_0^{eV} d\epsilon \sum_{\{n_l\}} \vert C_{\{n_l\}} \vert^2\delta 
(\epsilon-E_{\{n_l\}}), \label{IVChar} \eea
which is essentially the sum over all the squared spectral weights of states 
with excitation energy $\epsilon<eV$ (Ref. [\onlinecite{zuelicke96}]). 

In Fig. \ref{IVResultsN36}, we show the $I-V$ characteristics computed for a 
system of $N=75$ particles.
Log-log plots in these panels show several plateaus and steps in the low voltage region, which are purely due to the finite size effect of summing over a discrete set of spectral weights (in the low-energy regime we have very few states in spite of the fairly large number of particles considered). We observe, surprisingly in view of the physics described in the introduction, that 
the exponent $\alpha$ in $I\sim V^\alpha$ remains very close to the ideal EFTA result of 3, within numerical errors. To explore the reasons behind the robustness of the edge exponent to nonlinearities in the dispersion, we have plotted the energy at the maxima of the spectral function $A(k,\epsilon)$ as a function of $k$ in panel (d) of Fig. \ref{SFFigN36}. In the energy region of interest ($\epsilon \le 0.2$), the peaks roughly follow the ideal EFTA line. The low-energy states near the lower edge of the dispersion have comparatively less spectral weight, and their contribution to the tunneling density of states is negligible. 

\subsection{Irrelevance of edge roton in tunneling}

\begin{figure*}[htbp]
\begin{minipage}{1.0 \linewidth}
\includegraphics[scale=1.0,viewport=60 610 575 790,clip]{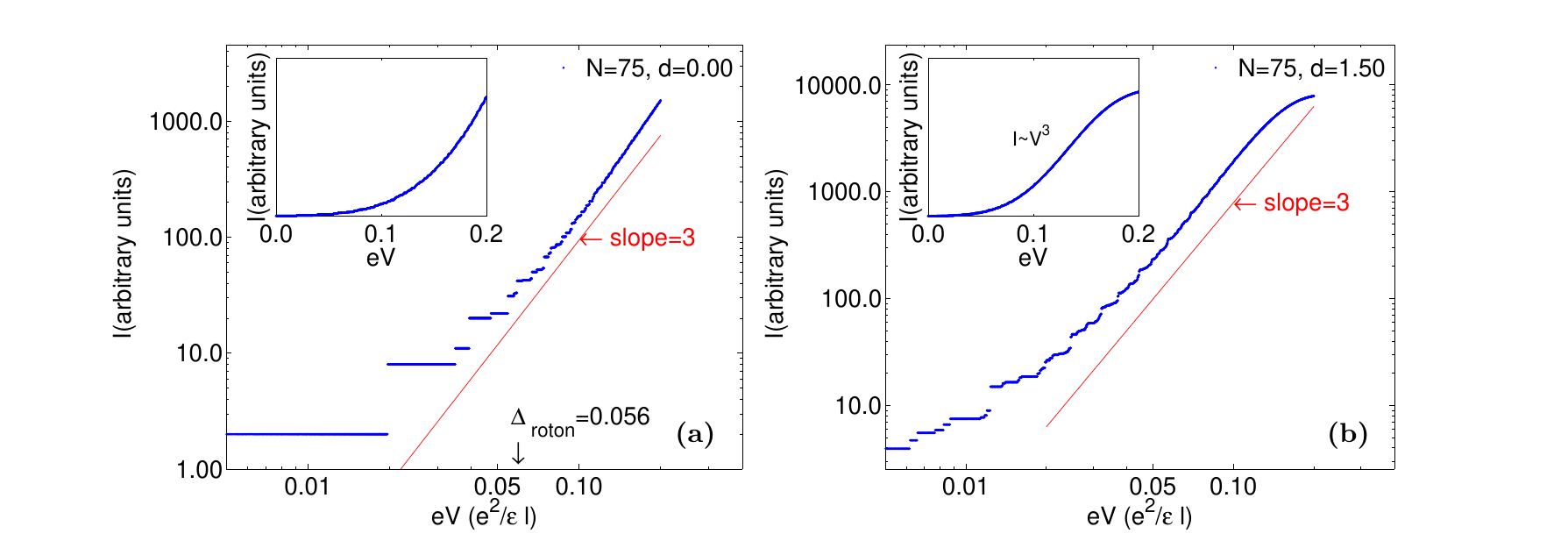}
\end{minipage}
\caption{(Color online) (a) The inset shows $I$ as a function of $V$ for tunnel transport between a FQH edge and 
a Fermi liquid computed from Eq. (\ref{IVChar}) using the numerical edge dispersion for $N=75$ particles. The main panel shows the log-log plot of $I-V$ characteristics, to better bring out the power law behavior. The red line marks the curve $I\propto V^3$. The step like deviations at small energies are a finite size artifact (see text for explanation). A logarithmically small deviation can be seen at $eV=\Delta_{\rm roton}$, the position of which has been marked. (b) Same as in panel (a) but for a setback distance where the roton gap $\Delta$ vanishes. The resulting slope is still close to 3 at energies $\epsilon <0.15$.}
\label{IVResultsN36}
\end{figure*}

One might ask whether the edge roton produces any signature 
in a tunneling experiment. In panel (a) of Fig. \ref{IVResultsN36}, no significant structure is seen when $eV$ is equal to the roton energy. Panel (b) corresponds to the setback distance where the roton gap just vanishes. Again, there is no prominently visible structure that may be attributed to the roton energy. An increase in the density of states at very low energies, shown in the log-log plots, is associated with the edge roton, but such a signature would be difficult to detect in experiments. We surmise that the spectral weight in the roton mode is too small for it to be observable in tunnel transport.

\section{Effective approach for reconstructed edge and tunneling exponent}
\label{sec:FieldEdRec}

For systems which undergo edge reconstruction (in our case $d>d_c$), a logical 
procedure would be to study the excitations around the {\em new} ground state
which now occurs at a finite $\Delta M$. This, however, is not possible 
in our numerical calculations because of computational limitations. For 
example, in the 27 or 45 particle system, we cannot go to large enough values 
of $\Delta M$ to identify the minimum energy.

To make further progress, we make the assumption that the 
edge-reconstructed system can be described by multiple chiral edges (we take 
three chiral edges below), which interact with one another. For want of a 
better description, we further assume that each chiral edge can be modeled by 
the EFTA Lagrangian and ask to what extent this can describe the experiments.

We will use the technique of bosonization to study the effects of 
density-density interactions between three chiral modes. A model with three 
modes (two of them moving in one direction and the other mode moving in the 
opposite direction) is motivated by the analysis in Refs. 
[\onlinecite{Yang03,Orgad}]. We denote the modes as 1, 2 and 3, of which 1 and 
3 move to the right (from $x=-\infty$ to $\infty$) and 2 moves to the left 
(from $x=\infty$ to $-\infty$). We take the bosonic Lagrangian density for 
the system to be of the EFTA form (we use a slightly different normalization 
for the bosonic field in this section than in Sec. V),
\beq {\cal L} = - \frac{1}{4\pi} \int_{-\infty}^\infty dx[\sum_{p=1}^3
\ep_p \pa_t \phi_p \pa_x \phi_p + \sum_{p,q=1}^3 \pa_x \phi_p K_{pq} 
\pa_x \phi_q ], \label{lag} \eeq
where $\ep_p = 1$ and $-1$ for the right and left moving fields respectively,
and $K_{pq}$ is a real symmetric matrix whose off-diagonal entries give the
strengths of the interactions between pairs of modes; we have assumed these
interactions to be short-ranged for simplicity. We have absorbed the 
velocities $v_p$ of the modes in 
the diagonal parameters $K_{pp}$. For repulsive density-density interactions,
the diagonal elements $K_{pp}$ as well as $K_{13}$ are positive,
while $K_{12}$ and $K_{23}$ are negative; this is because the densities of
fields 1 and 3 are given by $(1/2\pi) \pa_x \phi_1$ and $(1/2\pi) \pa_x 
\phi_3$, while the density of field 2 is $-(1/2\pi) \pa_x \phi_2$. Note that 
the filling factors $\nu_p$ ($p=1,2,3$) have not been introduced in the 
Lagrangian density in Eq. (\ref{lag}). They will appear later when we consider
the electron creation operator on edge $p$, namely, $\psi_p \sim 
e^{i \phi_p/\sqrt{\nu_p}}$.

To quantize the above theory, we impose the equal-time commutation relations 
$[\phi_p (x), \rho_q (y)] = -i \de_{pq} \de (x-y)$. These are satisfied if 
the fields have the decomposition at time $t=0$, 
\beq \phi_p (x) = \int_0^\infty \frac{dk}{k} [b_{pk} e^{i \ep_p kx} +
b_{pk}^\dag e^{-i \ep_p kx}], \eeq
where $[b_{pk}, b_{qk'}^\dag ] = \de_{pq} ~k ~\de (k-k')$.

In the absence of off-diagonal interactions (i.e., $K_{pq} = 0$ if $p \ne q$),
the velocities of the three modes are given by $K_{11}$, $-K_{22}$ and 
$K_{33}$; the first and third are positive, while the second one is negative.
When the off-diagonal interactions are present, the Lagrangian density in Eq. 
(\ref{lag}) can be diagonalized either by a Bogoliubov transformation 
[\onlinecite{tsallis}] or, equivalently, by solving the equations of motion. 
In the latter method, we assume that the fields take the form
$\phi_p = X_{p\al} e^{ik(x - \tv_\al t)}$, where the index $\al$ ($=
1,2,3$) labels the three different solutions, and $\tv_\al$ denote the 
corresponding velocities. The equations of motion then show that the 
eigenvectors $X_{p\al}$ (which are real) and the velocities $\tv_\al$ are 
solutions of the equations
\beq \sum_{q=1}^3 \ep_p K_{pq} X_{q\al} = \tv_\al X_{p\al}. \label{eigen} \eeq
We again assume that the new velocities $\tv_\al$ are positive for $\al =1,3$
and negative for $\al =2$, and we define $\ep_\al = 1$ for $p= 1,3$
and $-1$ for $\al = 2$. The eigenvectors $X_{p\al}$ can be normalized so that
\bea \sum_{p=1}^3 \ep_p \ep_\al X_{p\al} X_{p\be} &=& \de_{\al \be}, \non \\
\sum_{\al=1}^3 \ep_p \ep_\al X_{p\al} X_{q\al} &=& \de_{pq}. \label{norm} \eea
If $b_{pk}$ and $\tb_{\al k}$ denote the original and new (Bogoliubov 
transformed) bosonic annihilation operators, we find that these are related as
\bea \tb_{\al k} &=& \sum_{p=1}^3 X_{p\al} [ \frac{1}{2} (1 + \ep_p
\ep_\al) b_{pk} - \frac{1}{2} (1 -\ep_p \ep_\al) b_{pk}^\dag ], \non \\
b_{pk} &=& \sum_{\al=1}^3 X_{p\al} [\frac{1}{2} (1 +\ep_p \ep_\al)
\tb_{\al k} + \frac{1}{2} (1 - \ep_p \ep_\al) \tb_{\al k}^\dag ]. \non \\
&& \label{transf} \eea

Using Eq. (\ref{norm}), we can verify that $[b_{pk}, b_{qk'}] =0$ and 
$[b_{pk}, b_{qk'}^\dag ] = \de_{pq} ~k ~\de (k-k')$ imply that $[\tb_{\al k},
\tb_{\be k'}] =0$ and $[\tb_{\al k}, \tb_{\be k'}^\dag ] = \de_{\al \be} ~k ~
\de (k-k')$, as desired.

Let us now consider the electron creation operator on one of the three edges, 
say, $\psi_1 \sim e^{i\phi_1 /\sqrt{\nu_1}}$, where we have assumed
that edge 1 is associated with the filling factor $\nu_1$. In the absence of
the off-diagonal interactions, $\psi_1$ has the scaling dimension $1/(2\nu_1)$.
In the presence of interactions, we find from Eq. (\ref{transf}) that the 
scaling dimension of $\psi_1$ is given by
\beq d_1 ~=~ \frac{(X_{11})^2 + (X_{12})^2 + (X_{13})^2}{2\nu_1}. \label{scale}
\eeq
Since the second equation in Eq. (\ref{norm}) with $p=q=1$ implies that 
$(X_{11})^2 - (X_{12})^2 + (X_{13})^2 =1$, the expression in 
Eq. (\ref{scale}) is larger than $1/(2\nu_1)$ if $X_{12} \ne 0$, i.e., if 
there is a non-zero interaction $K_{12}$ between modes 1 and 2. We thus see 
that interactions between two counter-propagating modes lead to an increase 
in the scaling dimension of the electron operator; hence the exponent for the 
two-point correlation function for electrons becomes larger than $1/\nu_1$. 
In particular, if an edge corresponds to a filling factor of $1/3$ or less, 
the electron correlation exponent on that edge will be larger than 3. Thus a 
model with multiple chiral modes in which counter-propagating edges interact 
with each other has difficulty in explaining the results of tunneling 
experiments [\onlinecite{Chang3,Grayson,Chang1,Chang2}] which measure an 
exponent of about $2.7$. 

Wan {\em et al} [\onlinecite{Wan03}] and Joglekar {\em et al} 
[\onlinecite{Joglekar}] also studied the reconstruction of FQH edges at 
$\nu =1/3$ and showed that the presence of counter-propagating edges leads to 
a non-universal exponent. Yang [\onlinecite{Yang03}] introduced an action 
which has cubic and 
quartic terms in bosonic fields and showed that this leads to an exponent 
slightly larger than 3. In contrast, we have considered a standard action 
which is quadratic in bosonic fields and have shown that interactions between 
counter-propagating modes necessarily leads to an exponent larger than 3.

\section{Discussion and Conclusions}

We have investigated the influence of nonlinear dispersion on the 
physics of the FQH edge at $\nu=1/3$. Our approach involves microscopic 
calculations of the edge dispersion and the associated bosonic spectra, and 
the use of spectral weights from the bosonic theory. 

The conclusions of our work are as follows.

\begin{enumerate}

\item The edge dispersion is linear for energies below 
$0.02-0.04 e^2/\epsilon l$ ($0.2$meV to $0.4$meV) depending on the 
electron-background separation. For $d<d_c=1.5l$, an edge magnetoroton is 
observed. The maximum roton gap is $\Delta \approx 0.056 (e^2/\epsilon l)$ 
for zero setback distance.

\item Edge reconstruction occurs beyond a critical electron-background 
separation $d_c\approx 1.5l$ for smooth edges of a $\nu=1/3$ system, in 
agreement with the previous literature [\onlinecite{Wan03}]. 

\item A bosonic description of the edge excitation spectrum is satisfactory.
It requires the dispersion of the single boson excitation as an input.

\item The spectral weights of the electronic dispersion, though individually 
different from that of predictions of the bosonic theory, obey the same sum 
rules for a given angular momentum (provided $\Lambda$ level mixing is 
neglected). 

\item The tunneling exponent is surprisingly insensitive to the nonlinearity 
in the edge boson dispersion. The peaks of the spectral function for 
different momenta roughly follow the linearity of the ideal EFTA. The 
low-energy states have a small spectral weights and contribute negligibly 
to the tunneling. 

\item The roton has no significant contribution to the spectral function and 
hence to the tunneling density of states. Only a logarithmically weak 
signature of the roton may be observed in tunneling experiments. 

\item It is well known that the model assuming a single chiral mode is not 
adequate for understanding the results of experiments on systems which 
undergo edge reconstruction. An effective theory description with three 
chiral edges at $\nu=1/3$ produces an exponent that is larger than 3, 
contrary to the experimental finding of a smaller-than-3 exponent. 

\end{enumerate}

\section{Acknowledgments}

We acknowledge Paul Lammert, Chuntai Shi, Sreejith Ganesh Jaya, and Vikas 
Argod for insightful discussions, support with numerical codes and cluster 
computing. The computational work was done on the LION-XC/XO and {\it Hammer} 
cluster of the High Performance Computing (HPC) group, The Pennsylvania State 
University.

\section{Appendix}

\subsection{Sum rules}
\label{sec:SUMrules}

The derivation of the sum rules in Eq. (\ref{SSWGenl}) for squared spectral weights at a given angular momenta $\sum_{\{n_l\}}ln_l= M$ is given below. 
Consider the multinomial expansion (Ref. [\onlinecite{Abramowitz}] p. 823)
\bea \left(\sum_{k=1}^\infty\frac{x_kt^k}{k}\right)^m &=& m!\sum_{n=m}^\infty 
\sum_{\{a_j\}} \frac{t^n}{n!}\frac{n!\prod_ix_i^{a_i}}{\prod_j (a_j!j^{a_j})},
\non \\
\sum_j j a_j &=& n, \eea
With the following transformations 
\bea m && \rightarrow b \quad \textrm{number of bosons}, \non \\
x_k && \rightarrow m \quad \textrm{inverse filling factor}, \non \\
a_j && \rightarrow n_j \quad \textrm{bosons occupation}, \non \\ 
n && \rightarrow M \quad \textrm{angular momentum}, \non \eea
we obtain
\bea \left(\sum_{k=1}^\infty\frac{mt^k}{k}\right)^b &=& b!\sum_{M=b}^\infty 
\sum_{\{n_j\}} \frac{t^M}{M!}\frac{M!\prod_i m^{ n_i}}{\prod_j (n_j!j^{n_j})},
\non \\
\sum_j j n_j &=& M. \eea
Hence we get
\beq \sum_{b=0}^\infty \left(\frac{1}{b!}\sum_{k=1}^\infty\frac{mt^k}{k}
\right)^b = \sum_{b=0}^\infty\sum_{M=b}^\infty \sum_{\{n_j\}}t^M
\prod_j\frac{m^{ n_i}}{n_j!j^{n_j}} \eeq
We simplify this by noting the relation
\bea \exp\left(m\sum_{k=1}^\infty\frac{t^k}{k}\right) &=& e^{-m\ln(1-t)} 
= \frac{1}{(1-t)^m}. \eea
The sum of the squared spectral weights [\onlinecite{PalaciosMacDonald}] is the coefficient of 
$t^M$
\bea 
\sum_{\{n_l\}}|C^{(m)}_{\{n_l\}}|^2 &=& \frac{(M+m-1)!}{M!(m-1)!}, \quad \sum_{l} ln_l = M. 
\label{Sumrule2} \eea

\subsection{Green's function}

The ideal EFTA assumes a linear dispersion $\epsilon(k)=v_Fk$, and the sum rule in Eq. (\ref{Sumrule2}). 
The Green's function for the 1D chiral edge is 
\bea G(x,t) &=& \langle 0 \vert T(\Psi(x,t)\Psi^\dagger (0,0))\vert 0\rangle 
\non \\
&=&\langle 0 \vert e^{-iHt} e^{ik0}\Psi(0,0)e^{-ikx}e^{iHt}\Psi^\dagger(0,0)\vert 0
\rangle, ~t>0. \non \\
&& \eea
We map the edge to a disk by setting $x=R\theta$ with radius $R=1$ and insert a complete set of states within the subspace of single boson modes: 
\beq
\sum_M \sum_{\{n_l\}}\vert M,\{n_l\}\rangle \langle M,\{n_l\} \vert =I_s ; \quad M=\sum_l ln_l.
\eeq
We make the following substitutions 
\bea k&=&\lambda M; \quad \lambda=M/\sqrt{6(N-1)} \non \\
\epsilon_{\{n_l\}}&=&\sum_l \epsilon_l n_l=\lambda v_F\sum_l ln_l=\lambda v_F M,
\eea
and proceed to calculate the Green's function: 
\begin{eqnarray}
G(x,t)&=&\sum_M \sum_{\{n_l\}} e^{-ikx}\langle 0 \vert \Psi(0,0)e^{iHt}
\vert M,\{n_l\}\rangle \non \\
&& \quad \quad \quad \times\langle M,\{n_l\}\vert \Psi^\dagger(0,0)\vert 0\rangle
\non \\
&=& \sum_M \sum_{\{n_l\}} e^{-i\lambda M (x-v_F t)} \langle M,\{n_l\} \vert
\Psi^\dagger(0,0)\vert 0\rangle \vert^2 \non \\
&=& \sum_M e^{-i\lambda M (x-v_F t)} \sum_{\{n_l\}} \vert \langle M,\{n_l\} \vert
\Psi^\dagger(0,0)\vert 0\rangle \vert^2 \non \\
&=&\sum_M e^{-i\lambda M (x-v_F t)} \binom{M+m-1}{m-1} 
\non \\
&\sim& \frac{1}{(x-v_Ft)^m}. 
\end{eqnarray} 
This shows how the power law follows from a combination of the linear dispersion and the sum rule.
For a general dispersion $\omega_k$, we evaluate the commutators of the 
bosonic fields to find the Green's function for the edge:
\bea G(x,t) &=& \frac{2\pi}{L}e^{m \sum_{k} \frac{1}{k}e^{-i(kx-
\omega_k t)}e^{-ak}}. \eea

\end{document}